# Macroscale three-dimensional proximity effect in disordered normal/superconductor nanocomposites

Katsuya Ueno[1], Nobuhito Kokubo[2], Satoru Okayasu[3], Tsutomu Nojima[4], Yukihito Nagashima[5], Yusuke Seto[6], Megumi Matsumoto[7], Takahiro Sakurai[7], Hithoshi Ohta[8], Kazuyuki Takahashi[1] & Takashi Uchino[1*]

[1]Department of Chemistry, Graduate School of Science, Kobe University, Nada, Kobe 657-8501, Japan

[2]Department of Engineering Science, University of Electro-Communications, Chofu, Tokyo 182-8585, Japan

[3]Advanced Science Research Center, Japan Atomic Energy Agency, Tokai, Ibaraki 319-1195, Japan

[4]Institute for Materials Research, Tohoku University, Sendai 980-8577, Japan

[5]Nippon Sheet Glass Co., LTD., Konoike, Itami 664-8520, Japan

[6]Department of Planetology, Graduate School of Science, Kobe University, Nada, Kobe 657-8501, Japan

[7]Center for Support to Research and Education Activities, Kobe University, Nada, Kobe 657-8501, Japan

[8]Molecular Photoscience Research Center, Kobe University, Nada, Kobe 657-8501, Japan

*e-mail:uchino@kobe-u.ac.jp

Proximity-induced superconductivity in superconductor (S)-normal conductor (N) junctions has motivated a number of theoretical and experimental studies, since it leads to genuine superconducting states in the normal region[1,2]. Recently, interest in S-N interfaces was renewed by the observation of exotic proximity effects in various systems, including S/semiconductor[3], S/ferromagnet[4], S/topological insulator[5], and S/graphene[6] structures. In general, the proximity effect is enhanced for transparent interfaces where coherent Andreev reflection is possible[7,8]. Also, it is generally believed that the proximity effect is, by definition, a localized phenomenon that can only be active in each S/N interface region. However, here we show that a three-dimensional (3-D) macroscale proximity effect is realized in a few-µm-thick $MgO/Mg_2Si/MgB_2$ nanocomposite layers with atomically smooth and clean heterointerfaces. We found from scanning superconducting quantum interference device (SQUID) microscopy measurements that the surface of the layer of more than 100×100 µm$^2$ undergoes transition into a bulk-like superconducting state although the layer contains only less than ~10 vol % of superconducting $MgB_2$ nanograins in a dispersed manner. In the proximity-induced superconducting region, vortex formation and annihilation processes as well as vortex-free Meissner regions were observed with respect to **applied fields in a similar manner as Abrikosov vortices in type-II superconductors. Furthermore, we found that the induced superconducting layers exhibit an anisotropic magnetization behavior, in consistent with the formation of the 3-D superconducting coherence. This unusually extended proximity effect suggests that disorder-induced interaction of Andreev bound states, which are coherent superposition of time reversed electron hole pairs, are realized in the nanocomposite. Thus, the present result not only expands the limit of the proximity effect to bulk scales, but also provides a new route to obtain a proximity-induced superconducting state from disordered systems.**

The spontaneous emergence of macroscopic phase coherence is one of the most intriguing phenomena in physics, which includes Bose-Einstein condensation, superfluidity, superconductivity and lasing. In conventional lasers, emission is stimulated into well-defined resonant modes of a Fabry-Pérot (FP) optical cavity, resulting in a coherent light beam (Fig. 1a). In analogy with a FP cavity, a well-designed S-N-S hybrid structure with atomically clean interfaces can also behave as an resonator[9], where electrons with energies below the superconducting gap are reflected as their time-reversed particles (holes)[10] to yield the resonant standing waves called Andreev bound states (ABSs) (Fig. 1b). Recently, ABSs have been recognized to play a key role in mesoscopic superconductivity and provide a underlying framework for Josephson effects and related novel quantum functionality[11-13].

Note, however, that the required feedback for laser action is provided also by light scattering of particles. In a strongly scattering medium, light can be trapped by recurrent multiple scattering events, leading to the formation of closed-loop paths and the related laser action called random laser[14-16] (Fig. 1c). Hence, one would expect that if appropriate in-phase conditions are satisfied, phase-coherent ABS-loops, which are electronic analogues to the closed-loop optical paths in random laser, can be created in a system where superconducting grains are randomly dispersed in a normal host (Fig. 1d). If these loops further interact in phase with each other as a whole, macroscale superconducting coherence could be introduced. However, realization of such loops, or even a single ABS, would be quite challenging because ideal Andreev reflection requires the S/N interface to be in the clean limit[17,18]. In usual disordered superconductors, such a clean condition will not be fulfilled because of the presence of various surface states and structural/electronic fluctuations[19,20].

Although the fabrication of high-quality heterointerfaces is rather a difficult task,

we have recently proposed a method to synthesize dense insulator(MgO)/ semiconductor($Mg_2Si$)/superconductor($MgB_2$) nanocomposites with atomically smooth interfaces by using solid phase reaction between metallic magnesium and a borosilicate glass[21]. The thus obtained nanocomposites exhibit a semiconductor-superconductor transition at ~37 K owing to the $MgB_2$ nanograins embedded in the MgO/$Mg_2Si$ matrix, eventually leading to the near-zero resistance state at 17 K. In contrast to the case of usual superconducting composites[22], percolation channels consisting of "physically" connecting superconducting grains do not exist in the MgO/$Mg_2Si$/$MgB_2$ nanocomposites because the volume fraction (~10 %) of $MgB_2$ in the normal host is far less than the percolation threshold (~50 %)[23]. These results allow us to assume that the phases of the superconducting wave function for various superconducting grains are locked together to give macroscale coherence due to the formation of the 3-D Josephson junction network. It has also been demonstrated by Chen and coworkers[24] that multiple Andreev reflection actually occurs in highly transparent $MgB_2$-MgO-$MgB_2$ trilayer Josephson junctions. Thus, the MgO/$Mg_2Si$/$MgB_2$ nanocomposite with clean interfaces will provide an ideal platform to investigate the expected ABSs-induced superconductivity in disordered systems. Hence, we here perform two different experiments to explore the possible macroscale proximity effect in the nanocomposite.

One is scanning SQUID microscopy, and the other is magnetic anisotropy measurements.

We prepared the $MgO/Mg_2Si/MgB_2$ nanocomposite by solid phase reaction between Mg and a sodium borosilicate glass, according to the procedure reported previously[21] (Methods). As reported in our previous work[21], periodic layers consisting of alternating a few micrometers-thick MgO- and $Mg_2Si$-rich layers are automatically fabricated (Fig. 2a). Selected area electron diffraction (SAED) measurements demonstrate that $MgB_2$ nanograins are present exclusively in the MgO-rich layers (Extended Data Fig. 1). The mole (volume) fractions of MgO, $Mg_2Si$, and $MgB_2$ in the MgO rich regions are estimated to be ~80 (~57), ~15 (~37) and ~5 (~6) % from X-ray photoelectron spectroscopy (XPS) elemental analysis (Extended Data Fig. 2). High-angle annular dark field-scanning transmission electron microscopy (HAADF-STEM) imaging of the MgO-rich layer (Fig. 2b) shows that $MgB_2$ nanograins with a typical size of ~20 nm, which are identified as dark black dots in the lower panel of Fig. 2b, are randomly dispersed with nearest-neighbour distances ranging from ~20 to ~50 nm. The high-resolution TEM (HREM) image of the MgO rich layer (Fig. 2c) revealed that the interface between the MgO and $MgB_2$ nanograins is rather continuous owing to the inplane alignment of $[11\bar{2}0]$ (0001) $MgB_2$ with $[1\bar{1}0]$ (111) $MgO^{25}$, showing no discrete grain boundaries and dislocations. We confirmed that the present nanocomposite exhibits a two-step superconducting transition at $T_{c1}$=37 K and $T_{c2}$=19 K (Extended Data Fig. 3), which are attributed to the intragrain and intergrain phase-locking superconducting transitions[21], respectively. The room temperature resistivity of the composite is rather low (~0.03 Ωcm), also demonstrating a good electrical contact in between the constituent grains.

When we properly polished the surface of the composite in the direction parallel to the reaction layers, we can obtain the region where the polished top surface consists mostly of MgO (Fig. 2d). Using the thus polished surface, we performed scanning SQUID microscopy (SSM) measurements. Figure 3a shows a typical SSM image of the polished surface of a 120 μm × 120 μm area obtained under a nominally zero-field cooled (ZFC) condition at 29 K, which is below $T_{c1}$ (37 K) but above $T_{c2}$ (19 K). The distribution of vertical magnetic flux density $B_z$, which results from a vertical component of the residual magnetic field in the measurement system, is almost homogeneous over the entire scanned region, yielding the average magnetic flux density of 3.2 μT (see also Supplementary Video 1). This indicates that the residual magnetic flux penetrates uniformly between the $MgB_2$ grains; that is, the host matrix is mostly in the normal state although the embedded $MgB_2$ grains might be in the superconducting state. However, when the temperature is lowered down to 4 K, which is well below $T_{c2}$, vortices-like features along with vortex-free Meissner regions are recognized in the SSM image (Fig. 3d, Supplementary Video 2). We then performed a series of SSM measurements under field-cooled (FC) conditions by applying positive and negative magnetic fields perpendicular to the surface plane. We found that the vortex-like patterns become more (less) apparent by applying more positive (negative) values of magnetic field (Fig. 3b–h, Supplementary Videos 3−6). Note also that the applied field of −3 µT was found to reduce the local magnetic flux density down to ~0 nT over the whole scanned region (Fig. 3h, Supplementary Video 6). Thus, the −3-µT SSM image practically represents the ZFC state of the induced superconducting phase (see Extended Data Fig. 4 for the SSM images obtained under −3 µT at different temperatures). It should also be worth mentioning that each bright spot shown in the upper panels in Fig. 3 indeed corresponds to a single vortex carrying one flux quantum ($\Phi_0$), which was confirmed by integrating the observed $B_z$ values around the spot. These observations allow us to conclude that at temperatures well below $T_{c2}$, the whole scanned region undergoes transition into a bulk-like superconducting state where magnetic field penetration is observed in the form of flux quanta (Josephson vortices) in much the same way as Abrikosov vortices nucleate in hard type-II superconductors. In general, direct imaging of Josephson vortices in SNS junctions has been quite challenging[26], and, to our knowledge, this is the first report showing the formation and annihilation processes of Josephson vortices in such macroscale N regions.

The above arguments are further confirmed by magnetic moment (*M*) measurements. In the temperature range from ~25 to 37 K, we obtained basically the same initial magnetization curves even when the direction of applied field *H* is changed from a perpendicular ($H\perp$) to an inplane ($H//$) direction along the reaction layers (Fig. 4a,b). When the temperature is lowered below ~20 K ($T < T_{c2}$), however, an additional (and much steeper) diamagnetic component emerges in the low *H* region, which results from the induced superconducting phase. The absolute initial slope is an order larger for $H\perp$ than for $H//$ at temperatures below ~20 K (Fig. 4c), indicating that the superconducting volume fraction under the $H\perp$ condition is about ten times larger than that under the $H//$ condition. The flux pinning anisotropy is observed in the *M*(*H*) loops as well (Extended Data Fig. 5). Also, temperature dependent magnetization under ZFC condition is more negative for $H\perp$ than for $H//$ at temperatures below ~20 K (Fig. 4d,e). We further found that the irreversible temperature $T_{irr}$, at which FC- and ZFC-*M*(*T*) curves begin to separate (see the inset in Fig. 4e), obeys a simple power law of the type, $H \propto [1- T_{irr}(H)/T_c(0)]^n$ ($T_c(0) = 37$ K), similar to the case of cuprate superconductors[27].

The power law coefficient *n* also exhibits anisotropy (Fig. 4f). These anisotropic features at temperatures below $T_{c2}$ are indicative of a change in dimensionality of the superconducting phase from nearly 0-D (particles) to 3-D (bulk). It is probable that the penetration depth in the induced superconducting region is on the order of micrometers, as in the case of the penetration depth in conventional Josephson junctions[28]. If the penetration depth is longer than or comparable to the thickness of the MgO-rich layers, the effect of field penetration on the magnetization will become more pronounced for $H//$ than for $H\perp$ as long as the whole MgO-rich layer behaves as a bulk superconductor.

Previously, an anomalously large proximity effect, extending over a length scale 100 times longer than the superconducting coherence length, has observed in atomically smooth S-N-S junctions where the component S is an optimally doped high-Tc cuprate and N is also of the same family, but in the underdoped normal state[29]. In that case, a pseudogap state in the S part has been proposed to be responsible for the anomalous proximity effect[30]. However, our observations imply that the interface states also play a key role in exhibiting the long-range proximity effect. Thus, our experiments will not only reveal interesting physics behind the behavior of Cooper pairs in disordered but clean superconductors, in view of the model shown in Fig. 1, but may provide a new material platform for future advanced quantum devices using Josephson vortices.

**METHODS**

**Sample preparation.** The MgO/Mg$_2$Si/MgB$_2$ nanocomposite used in this work was obtained by reacting metallic Mg with a sodium borosilicate glass under argon environment[21]. Since borosilicate glass includes SiO$_2$ and B$_2$O$_3$ as glass forming oxides, the resulting reactions can be represented as follows[31,32]:

$$4Mg + SiO_2 \rightarrow 2MgO + Mg_2Si, \qquad (1)$$

$$4Mg + B_2O_3 \rightarrow 3MgO + MgB_2. \qquad (2)$$

Sodium was added in the form of Na$_2$O, which facilitates stable glass formation. We[21] confirmed that Na$_2$O was completely expelled from the reaction zone during reaction between Mg and the sodium borosilicate glass. Hence, the reaction products consist practically of MgO, Mg$_2$Si and MgB$_2$.

The starting sodium borosilicate glass with a typical composition of 68SiO$_2$–24B$_2$O$_3$–8Na$_2$O (in mol %, B$_2$O$_3$/SiO$_2$=0.35) was prepared using a conventional melt-quenching method. When the B$_2$O$_3$/SiO$_2$ ratio of the starting glass exceeds over 0.4, the glass becomes softer and more fluid during reaction with Mg. This prevents the formation of well-aligned layered structures in the product, showing no two-step superconducting transition in the electrical resistivity (Extended Data Fig. 6).

On the other hand, a clear superconducting transition with near-zero resistivity was not observed when we used the glass with the $B_2O_3/SiO_2$ ratio is below ~0.3 (Extended Data Fig. 6), although the well-aligned layers were found to be formed in the product. The lack of a clear superconducting transition most likely results from the fact that the concentration of superconducting $MgB_2$ grains is too low to exhibit the expected 3-D proximity effect. Thus, the optimum $B_2O_3/SiO_2$ ratio of the starting glass composition for attaining desirable structural and electrical properties is ~0.35.

Commercial powders of $SiO_2$, $B_2O_3$, and $Na_2CO_3$ were mixed and melted in a platinum-rhodium crucible at 1400–1500 °C for 2 h. The melt was then poured onto a steel plate and was annealed at 600 °C for 24 h. The resulting glass was cut into cubic pieces with a size of 10×10×10 mm and polished with cerium oxide. Then, the cubic glass sample and ~1 g of Mg powders were put in a cylindrical alumna crucible with an internal diameter of 36 mm and a height of 27 mm. This crucible was located inside a larger rectangular (90×90×50 mm) alumina crucible, which was closed with a 4-mm thick alumina lid. This set of crucibles was placed in a box-type electric furnace. The furnace was evacuated to a pressure down to ~30 Pa and purged with argon. In order to promote the reaction between Mg and sodium borosilicate glass, the temperature of the furnace was raised to 700 °C at a rate of ~10 °C/min and kept for 5 h under flowing argon environment. After the heating process, the reaction layer with a thickness of ~0.5 mm was developed at all the six surfaces of the cubic glass sample. We then mechanically removed the reaction layer from the unreacted region of the glass. For the SSM measurements, both the top and bottom surfaces of the thus removed sample were polished in the direction parallel to the layered structures with a diamond paste with an average particle size of 250 mm. For the XRD and density measurements, the samples were further crushed into fine powder.

**Characterization.** Powder X-ray diffraction (XRD) patterns were obtained with an X-ray diffractometer (Rigaku, SmartLab) using Cu K$\alpha$ radiation (wavelength $\lambda$=1.5418 Å). The density of the sample was measured by a helium pycnometer (Shimadzu, AccuPyc II 1330). Scanning electron microscopy (SEM) and energy dispersive X-ray (EDX) spectroscopy were conducted with a scanning electron microscope (JEOL, JSM-5610VS) with an EDX spectrometer. X-ray photoelectron spectroscopy (XPS) was carried out with a XPS spectrometer (Ulvac-Phi, PHI X-tool) utilizing Al K$\alpha$ X-rays (1486.6 eV). High-resolution transmission electron microscopy (HR-TEM) measurements were performed with a (scanning) transmission electron microscope (JEOL, JEM-2100F) equipped with an EDX spectrometer and a high angle annular dark field detector (HAADF) operated at an acceleration voltage of 200 kV. An ion slicer was used to obtain a thin section for TEM measurements.

A four-terminal contact configuration was used for temperature-dependent resistivity measurements of the periodic layered structure. Four-terminal contacts were fabricated on the cross section of the product along the direction parallel to the layered structure. During the resistivity measurements, the temperature was varied from 4.2 to 300 K using liquid helium.

SSM measurements were conducted with a scanning SQUID microscope (SII Nanotechnology, SQM-2000) with a dc SQUID magnetometer consisting of Nb/Al-AlO$_x$/Nb Josephson junctions and an inductively coupled pick-up loop of a Nb film[33]. We used the pick-up loop of 10 μm diameter with 2 μm linewidth. The dc SQUID and the pick-up loop were integrated on a square Si chip of 3 mm×3 mm in size, which was mounted on a phosphor bronze cantilever. The sensor chip was tilted with respect to the sample stage by a shallow angle of ~10 degrees. A corner of the chip was slowly approached to the sample until a specific distance (~5 μm at the minimum) between the loop and the sample surface was achieved. Although the whole assembly including the sensor chip and the sample stage was surround by a μ-metal shield, a residual magnetic field (ambient field) of a few µT was present. A small magnetic field in the range of 0 – 3 µT was applied perpendicular to the sample surface using a coil wound in the sample stage. We scanned the sample stage in *XY* directions by dc stepper motors. During the measurements, the temperature of the SQUID sensor was kept to 2–3 K, whereas the sample was cooled to a desired temperature down to 4 K. The total spatial resolution of the present measurement system is ~ 5 µm.

A commercial SQUID magnetometer (Quantum Design, MPMS-XL) equipped with the reciprocating sample option (RSO) was used for magnetization measurements. Magnetization (*M*) in a ZFC states was measured by cooling the sample initially in a zero field to 2K, and ZFC magnetization was recorded in an applied magnetic field as the temperature is increased. When the temperature reached 300 K, the sample was gradually cooled under the applied filed to obtain the FC magnetization. The *M*(*H*) curves ware measured at several different temperatures after zero-field cooling.

**Data availability**

All data generated and analysed during this study are available within the paper and its Extended Data. Further data are available from the corresponding author on reasonable request.

**Supplementary Information**

1. Video 1: 3D animation of the SSM image shown in Fig. 3a.

2. Video 2: 3D animation of the SSM image shown in Fig. 3d.

3. Video 3: 3D animation of the SSM image shown in Fig. 3b.

4. Video 4: 3D animation of the SSM image shown in Fig. 3c.

5. Video 5: 3D animation of the SSM image shown in Fig. 3e.

6. Video 6: 3D animation of the SSM image shown in Fig. 3h.

**Acknowledgments**

The support of T. Yamada in the density measurements is acknowledged.

**Author Contributions**

Y. N. prepared the glass sample. K.U. fabricated the nanocomposites and carried out XRD, XPS and conductivity measurements. Y.S. and M.M. were in charge of the TEM imaging. N. K., S. O. and T. N. conducted SSM observations. T.U, T.S. and H.O. performed magnetic measurements. K.T. contributed with his experience and knowledge about the transport measurements. T.U. analyzed and interpreted the data, and wrote the manuscript with the assistance of all other co-authors.

**Author Information**

The authors declare no competing financial interests. Correspondence and requests for materials should be addressed to T. U. (uchino@kobe-u.ac.jp).

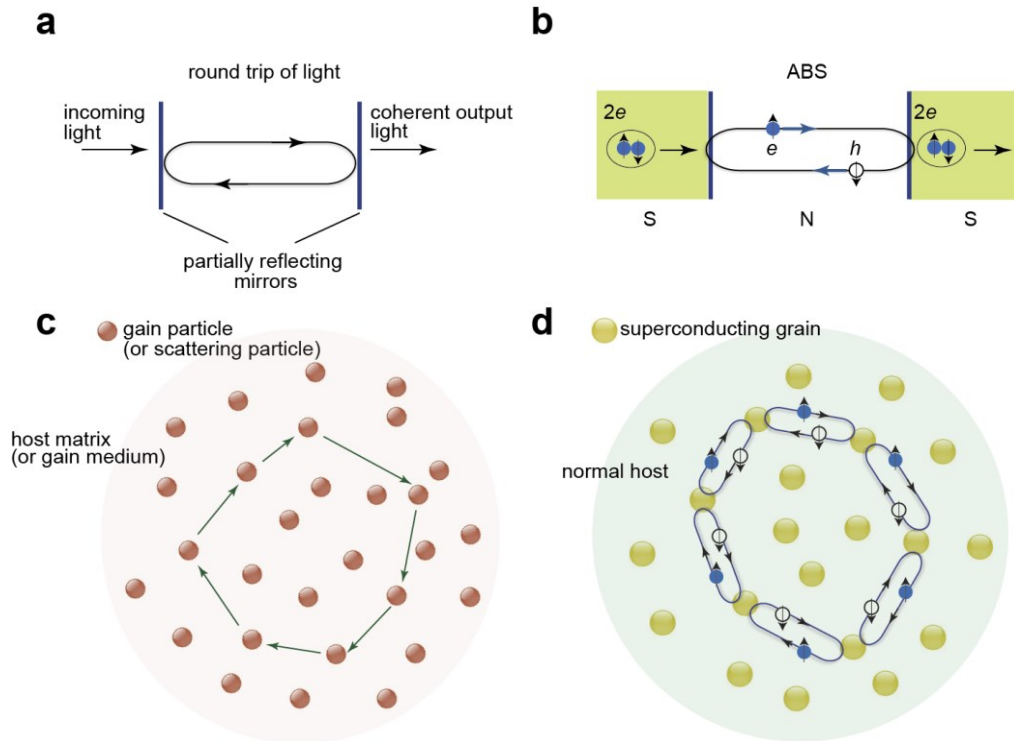

**Figure 1 | Comparison between optical and electronic resonators. a,** Schematic of a Fabry-Pérot cavity. **b,** Simplified schematic for Andreev bound state (ABS) formation in a S-N-S junction. When the energy of an electron (*e*) in the normal region is lower than the superconducting gap, the electron is reflected as a hole (*h*) in the opposite direction. The thus formed hole is reflected again as an electron, leading to the formation of resonant states of entangled *e*−*h* pairs, called ABSs. **c,** Working principle of random laser. Multiple scattering increases the path length of light inside the host matrix, resulting in a coherent loop for lasing. **d,** Schematic representation of the formation of a phase-coherent loop of ABSs. If the loop satisfies the condition for resonance, a long-range standing wave will be created in the loop. This loop is an electric analogue to that realized in random laser. A further spatial and coherent accumulation of these loops will eventually lead to a macroscale proximity effect.

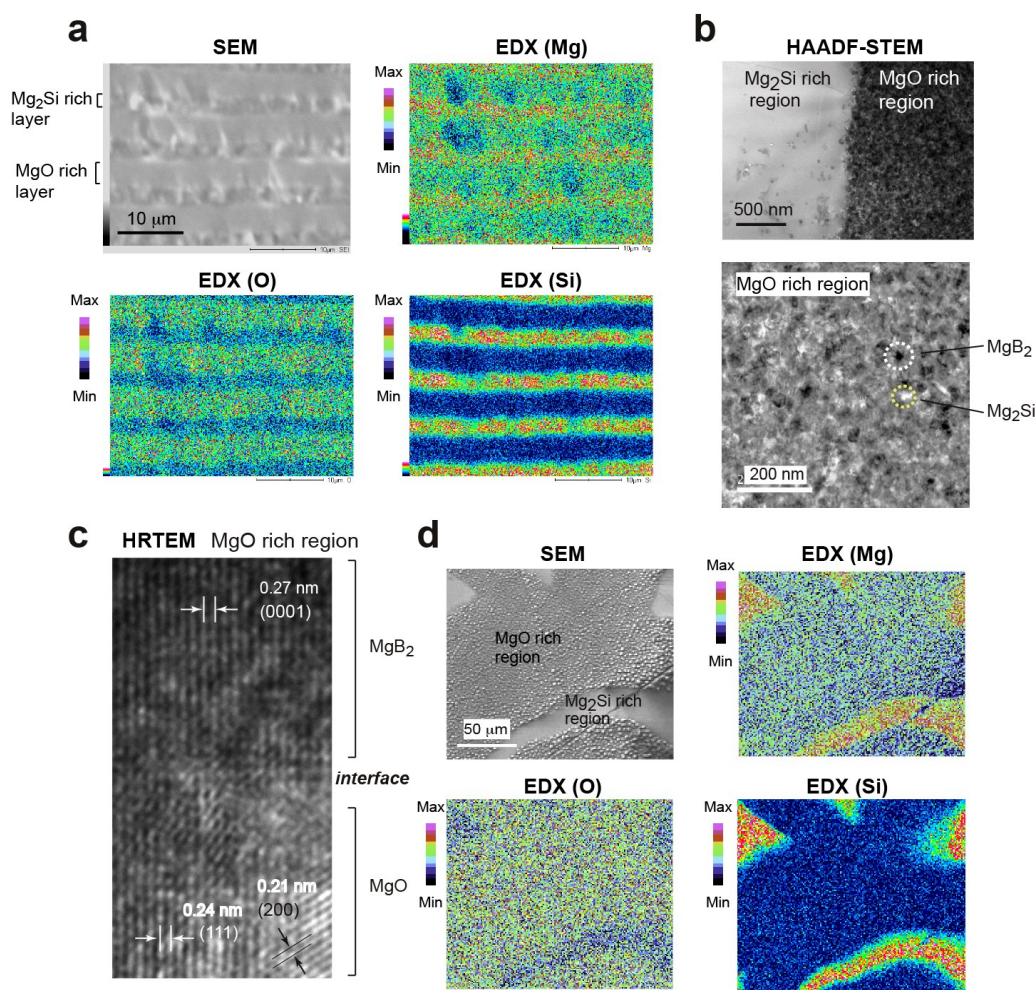

**Figure 2 | Electron microscope observations. a,** A SEM image of the cross-section of the reaction layers and the corresponding EDX elemental mapping images of Mg, O and Si. **b,** Low- (top panel) and high-magnified (bottom panel) HAADF-STEM images of the cross-section of the reaction zone. High-angle scattered electrons recorded using HAADF detector offer essentially an incoherent signal, and the intensity is approximately proportional to the square of the atomic number. This indicates that bright (dark) image contrast indicates heavy (light) elements. Thus, white, gray and dark black spots will correspond to the $Mg_2Si$, MgO and $MgB_2$ nanograins, respectively. **c,** An HRTEM image of the MgO-rich region showing the clean interface between MgO and $MgB_2$ nanocrystals. **d,** A typical SEM image of the polished top surface and the corresponding EDX elemental mapping images of Mg, O and Si.

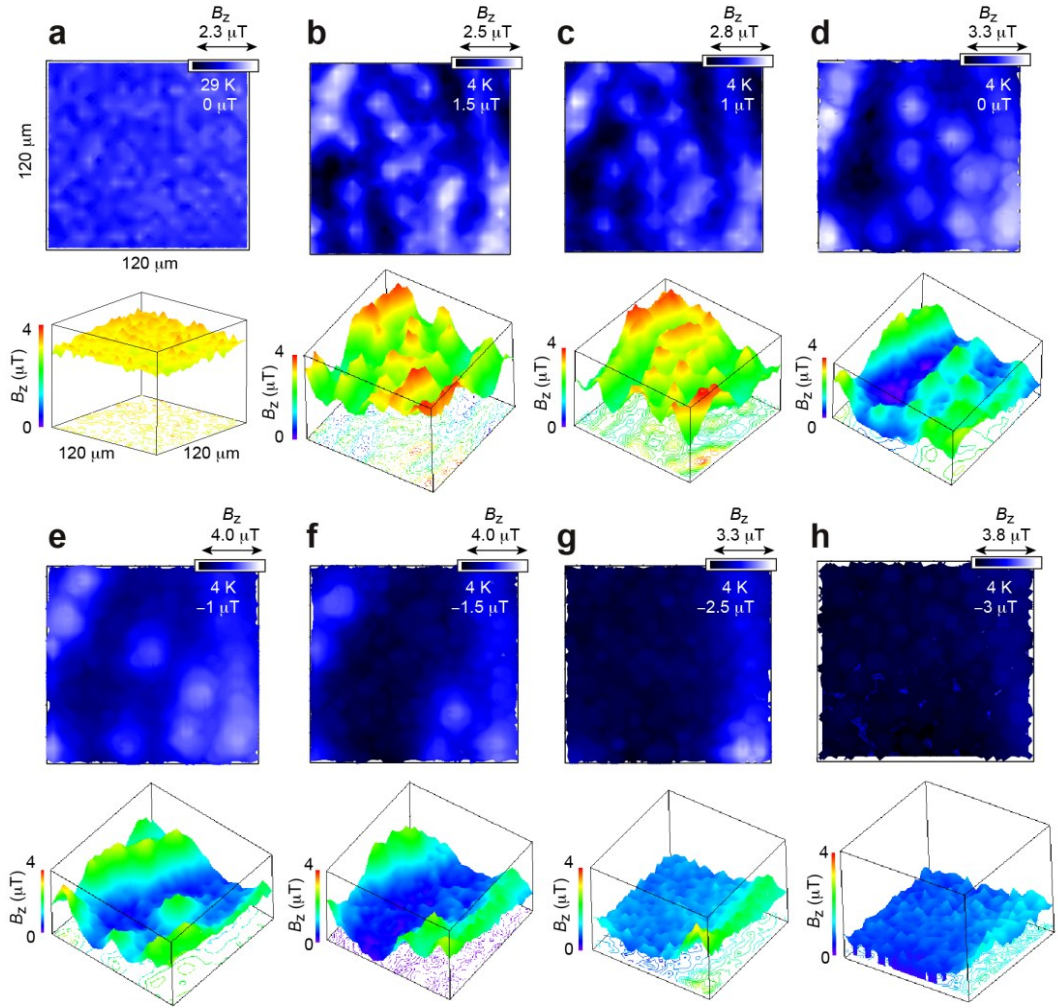

**Figure 3 | Scanning SQUID microscope images of the MgO-rich surface. a-h,** 2D (upper panels) and 3D (lower panels) representations of the vertical magnetic flux density $B_z$ for a 120 μm × 120 μm area obtained under different temperatures and applied fields: $T$ = 29 K at 0 μT (**a**), $T$ = 4 K at 1.5 μT (**b**), $T$ = 4 K at 1.0 μT (**c**), $T$ = 4 K at 0 μT (**d**), $T$ = 4 K at −1.0 μT (**e**), $T$ = 4 K at −1.5 μT (**f**), $T$ = 4 K at −2.5 μT (**g**) and $T$ = 4 K at −3.0 μT (**h**). Each bright spot shown in **b-f** corresponds to a single vortex carrying one flux quantum ($\Phi_0 = h/2e = 2.07 \times 10^{-15}$ Wb) since integration of the observed $B_z$ values around the spot yields the value of ~$2 \times 10^{-15}$ Wb.

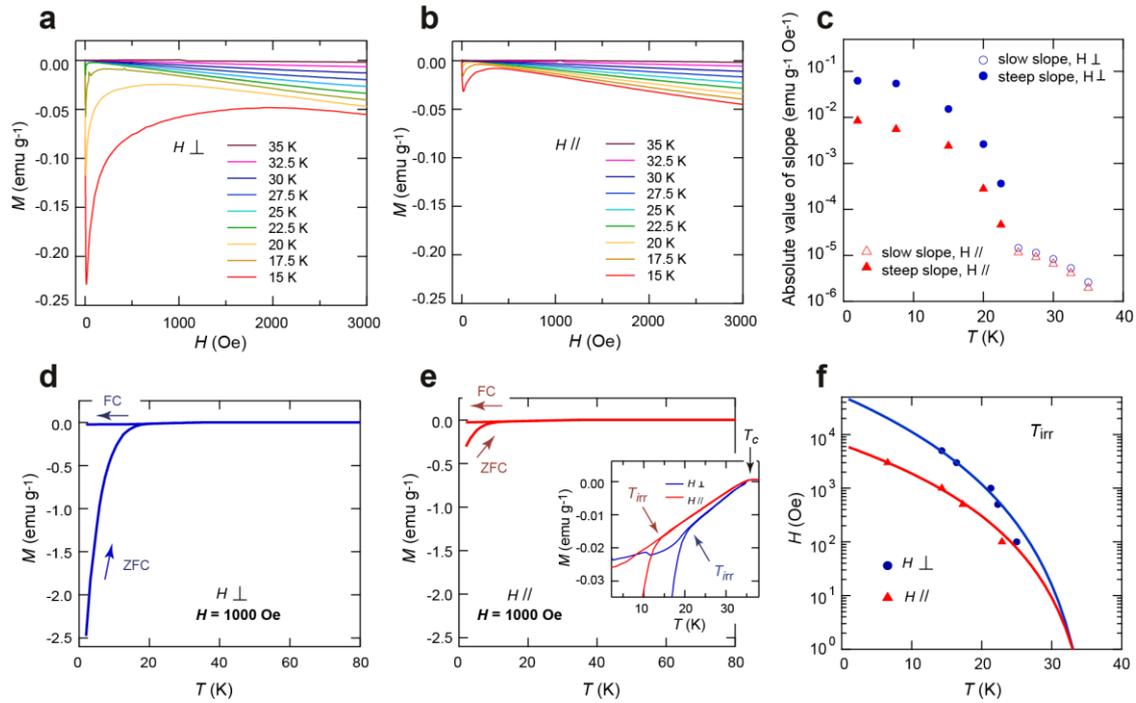

**Figure 4 | Magnetic anisotropy of the composite. a, b,** Initial $M(H)$ curves in the temperature region from 15 to 30 K measured under $H\perp$ (**a**) and $H//$ (**b**) conditions. **c,** Temperature dependence of the absolute value of initial slope in the $M(H)$ curves for $H\perp$ and $H//$ conditions. **d, e,** Temperature dependent magnetization curve under a field of 1000 Oe measured under $H\perp$ (**d**) and $H//$ (**e**) conditions. The inset in (**e**) shows a magnified view around the transition region for $H\perp$ and $H//$. The irreversibility temperature ($T_{irr}$) is defined as the point in which ZFC and FC curves separate. **f,** Irreversibility line obtained by performing ZFC and FC measurements under different applied fields from 100 to 3000 Oe and different field directions. The solid lines show the fitting result using the power low, $H \propto [1- T_{irr}(H)/T_c(0)]^n$ ($T_c(0)$ = 37 K). The fitted values of $n$ = 4.9 for $H\perp$ and $n$ = 3.9 for $H//$.

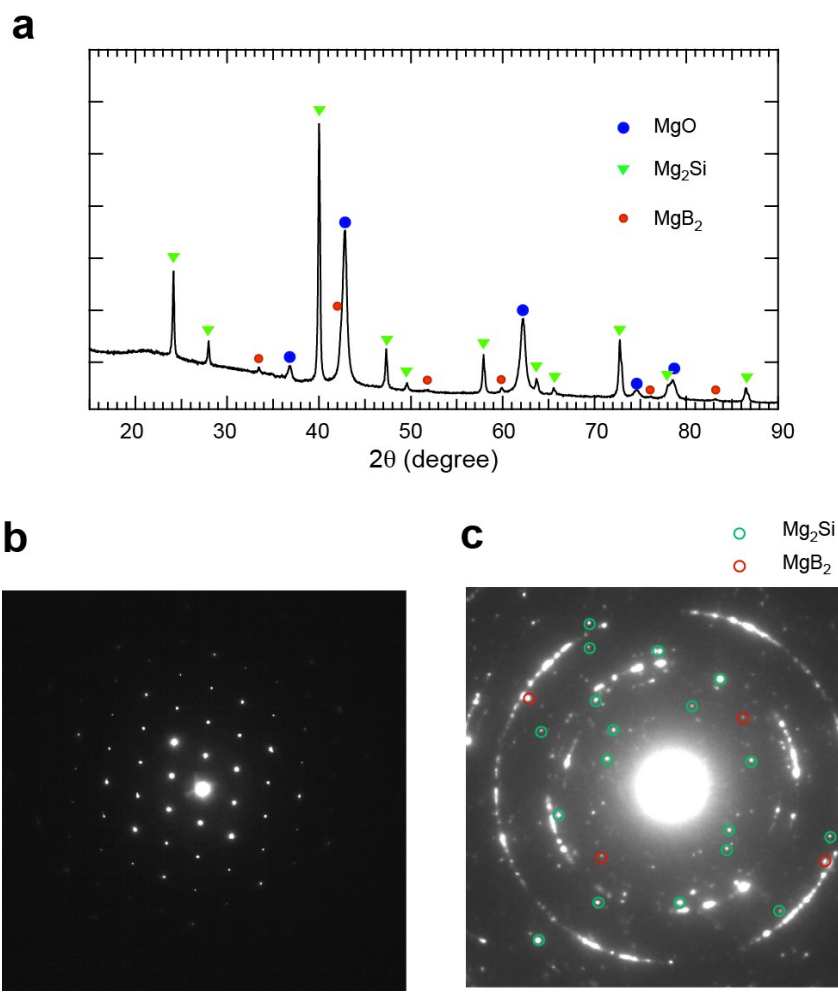

**Extended Data Figure 1 | X-ray and electron diffraction patterns. a,** The XRD pattern of the crushed powder taken with Cu Kα radiation (wavelength λ=1.5418 Å). Bragg peaks of MgO, $Mg_2Si$, and $MgB_2$ are shown. From the full-width at half-maximum of the X-ray Bragg peaks and the Scherrer formula, the sizes (lower limit) of MgO, $Mg_2Si$, and $MgB_2$ are estimated to be 30, 45, and 15 nm, respectively. **b,c,** SAED patterns of the $Mg_2Si$-rich layer (**b**) and the MgO-rich layer (**c**). The spotted diffraction pattern of the Mg2Si-rich layer (**b**) corresponds to the $[\bar{1}10]$ zone axis, indicating that the $Mg_2Si$-rich layer consists of a single crystalline-like $Mg_2Si$ phase. On the other hand, the MgO-rich layer (**c**) exhibits ring-like SAED patterns ascribed not only to MgO (major ring patterns) but also to $Mg_2Si$ (circled in green) and $MgB_2$ (circled in red).

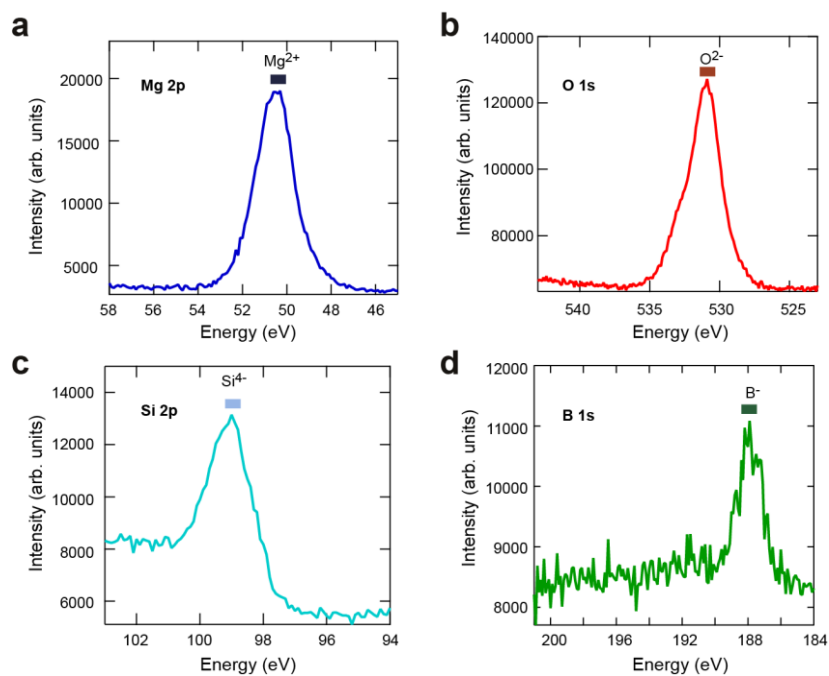

**Extended Data Figure 2 | X-ray photoelectron spectra (XPS) of the MgO-rich region. a,** Mg 2p, **b,** O 1s, **c,** Si 2p, and **d,** B 1s core level spectra. The sample was cleaned by repeated sputtering with inert ultrahigh pure Ar-gas in situ. The XPS spectra were referenced to the adventitious C 1s signal at 284.6 eV. The Si 2p peak at 99 eV in **c** is attributed to silicide anion in $Mg_2Si$ (Ref. 34), whereas the B 1s peak at 188 eV in **d** is attributed boride anion in $MgB_2$ (Ref. 35). The atomic concentrations estimated from the XPS elemental analysis are as follows: 40±8 % Mg, 39±8 % O, 7±1 % Si, and 4±1 % B. Accordingly, the mole (volume) percents of MgO, $Mg_2Si$, $MgB_2$ in the MgO-rich region are estimated to be ~80 (~57), ~15 (~37) and ~5 (~6) %, respectively.

If we assume that the glass with the starting composition of $68SiO_2–24B_2O_3–8Na_2O$ (mol. %) reacts fully with Mg via Eqs. (1) and (2) shown in Methods, and neglect the contribution from the sodium-related materials, the resulting average composition of the product is $69MgO-23Mg_2Si-8MgB_2$ (mol. %) (or $43MgO-49Mg_2Si-8MgB_2$ in vol. %). This estimated average composition yields the ideal density of 2.72 g/cm³, in good agreement with the observed sample density (2.70±0.3 g/cm³). Thus, the composition of the MgO-rich region estimated from the XPS elemental analysis is reasonable in that the fraction of MgO is substantially higher than that estimated from the average composition.

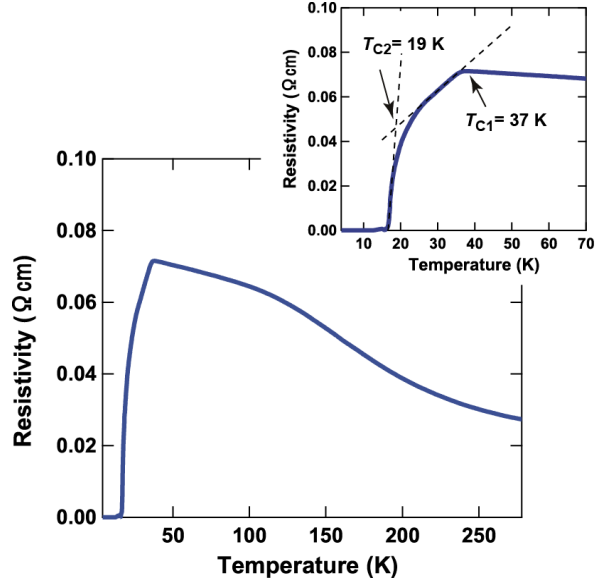

**Extended Data Figure 3 | Temperature dependence of electrical resistivity.** Electrical resistivity of the cross section of the product as a function of temperature along the direction parallel to the periodic layered structure. The inset shows a magnified view around the transition region, demonstrating a two-step transition. The higher and lower-transition temperatures are referred to as $T_{c1}$ and $T_{c2}$, respectively. $T_{c2}$ is defined as the point where extrapolations of the linear portions of the resistivity curve intersect. The room temperature resistivity is ~0.03 Ωcm, which is much lower than that (~5 Ωcm) reported in our previous paper[21]. We admit that the room temperature resistivity, or the quality and transparency of the heterointerfaces, varies from sample to sample. Note also that the present reactions are carried out under Mg-rich and O-deficient conditions in Ar atmosphere. Thus, a number of uncontrollable oxygen vacancies are expected to be present in the MgO nanograins. These oxygen vacancies can provide conduction channels via electron tunneling[36–38], further leading to a decrease in resistivity. However, the two-step superconducting transition along with the magnetic anisotropy shown in Fig. 4 is always observed irrespective of the resistivity value at room temperature. Thus, we consider that the macroscale proximity effect observed in this study is a generic phenomenon as far as the present nanocomposites are concerned.

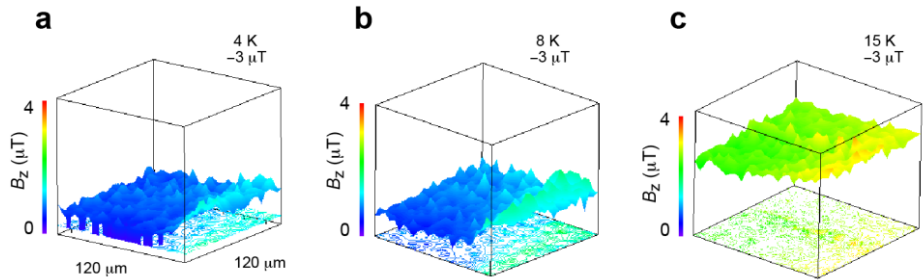

**Extended Data Figure 4 | SMM images obtained under −3 μT at different temperatures. a-c,** 3D representations of the vertical magnetic flux density $B_z$ for a 120 μm × 120 μm area obtained at $T$ = 4 K (**a**), $T$ = 8 K (**b**) and $T$ = 15 K (**c**) under field cooling of −3 μT. The scanned region is the same as that shown in Fig 3. The observed magnetic flux density increases on average with increasing temperature. This indicates that Meissner fraction in the induced superconducting region decreases with increasing temperature.

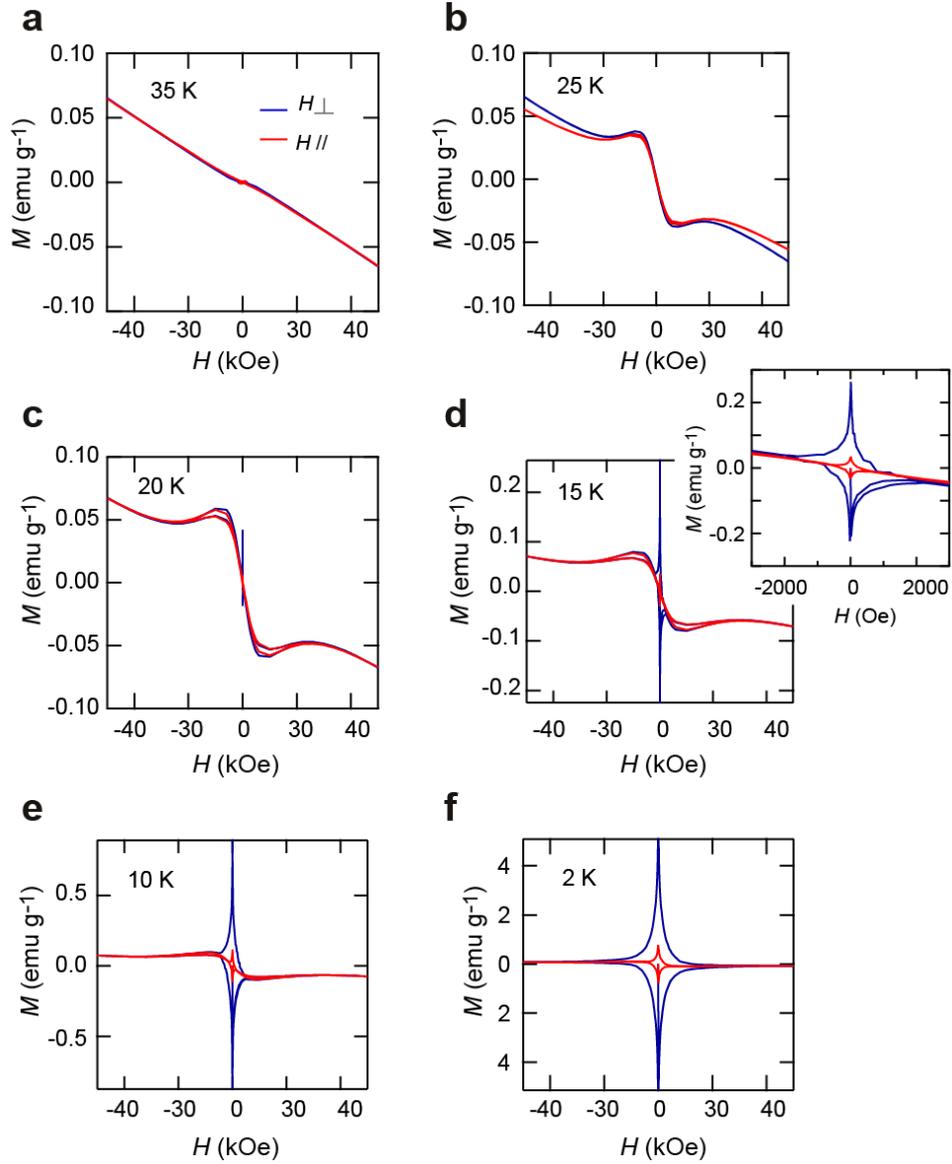

**Extended Data Figure 5 | Temperature dependent anisotropy of *M*(*H*) loops. a-f,** *M*(*H*) hysteresis loops measured for different field orientations H⊥ and *H*// at 35 K **(a)**, 25 K **(b)**, 20 K **(c)**, 15 K **(d)**, 10 K **(e)** and 2 K **(f)**. The inset in (**d**) shows a magnified view of the *M*(*H*) loops in the near-zero field region. At temperatures below 20 K, a wider hysteresis loop is observed for H⊥ than for *H*//, indicating that the pinning force is stronger for the vortices directed perpendicular to the MgO-rich layers than for those parallel to the MgO-rich layers. The observed flux pinning anisotropy is consistent with our assertion that the whole of the MgO-rich layers are transferred into the superconducting state at temperatures below ~20 K and behave as if it were a hard type-II superconductor with a penetration depth on the order of micrometers.

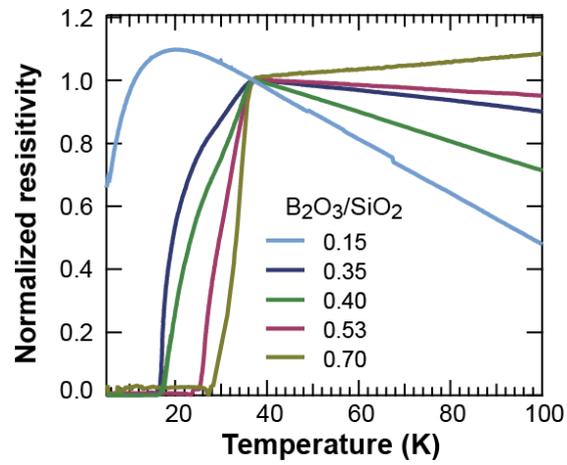

**Extended Data Figure 6 | Temperature dependence of the electrical resistivity of MgO/Mg$_2$Si/MgB$_2$ nanocomposites prepared from the sodium borosilicate glasses with different B$_2$O$_3$/SiO$_2$ ratios**. Electrical resistivity is normalized to its value at 37 K. The result for B$_2$O$_3$/SiO$_2$=0.35 corresponds to the one shown in Extended Data Fig. 3.